          [2004/02/13 1.1 (PWD)]
\documentclass[a4paper,twocolumn]{esapub} 

\usepackage{times}
\usepackage{natbib}
\usepackage{graphicx}

\title{Gamma-Ray All-Sky Imaging with BATSE}
\author{A. B. Hill}
\author{E. J. Barlow}
\author{A. J. Bird}
\author{A. J. Dean}
\author{C. Ferguson}
\author{S. E. Shaw}
\author{M. J. Westmore}
\author{D. R. Willis}
\affil{School of Physics and Astronomy, University of Southampton, Hampshire, SO17 1BJ, United Kingdom}

\begin{document}


\maketitle

\begin{abstract}
The BATSE mission aboard CGRO observed the whole sky for 9 years in
the 20 keV - 2 MeV energy band.  Flat-fielding of the
temporal variations in the background present in the data set has been
accomplished through a GEANT3 Monte-Carlo simulation - the BATSE Mass
Model (BAMM).  The Earth Occultation technique (EOT) is used together with a
maximum-likelihood imaging approach to construct all-sky maps with
$\sim$mCrab sensitivity.  Additionally, a non-linear CLEAN algorithm is
applied to the all-sky maps to remove large artefacts that are
systematic of the EOT.  The latest survey results produced through the
application of this technique to a subset of the BATSE data set are presented.
\end{abstract}

\section{Introduction}
BATSE was developed to detect and locate $\gamma$-ray bursts.  The
individual BATSE detectors have no positional sensitivity and hence the
position of a $\gamma$-ray burst was obtained by triangulation between
detectors.  However, before launch it was realised that persistent
$\gamma$-ray sources could be monitored by using the EOT
\citep{eot}.  As a source sets below the Earth's horizon a
downward step is seen in the detector count rates.  The height of a
step gives a direct measurement of the source flux.

The capability to generate an image has many advantages over simple
measurements of source fluxes.  In high energy astronomy all-sky
images have been produced between 0.1 - 2 keV by ROSAT \citep{rosat}
and $>$1 MeV by COMPTEL \citep{comptel}, however there is little
information regarding the intervening range, consisting almost
entirely of data from the HEAO-1 A4 experiment of 1978-79 \citep{heao}.
The 9 uninterupted years of all-sky monitoring by BATSE therefore
provides a wealth of additional information.   
 
\begin{figure}[htbp]
\centering
\includegraphics[width=1.0\linewidth, clip]{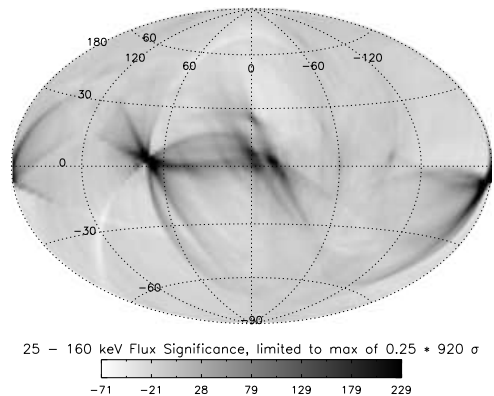}
\vspace{1cm}
\caption{Uncleaned map of the significance of the 25-160 keV flux from
the whole sky during the period TJD 09448-09936.\label{fig:raw_map}}
\end{figure}

\begin{figure*}[htbp]
\centering
\includegraphics[width=0.78\linewidth, clip]{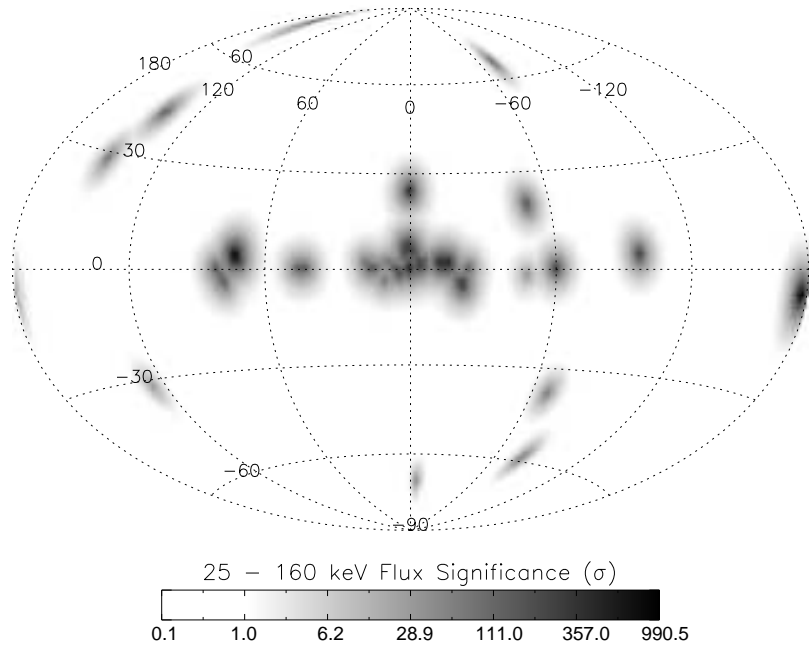}
\vspace{1cm}
\caption{The result of applying the CLEAN algorithm on the all-sky map
seen in Fig.~\ref{fig:raw_map}.  The sources are listed in Table~\ref{tab:cat}.\label{fig:clean_map}}
\end{figure*}

\section{Generating All-Sky Maps}
Mass Modelling is a technique to simulate the background radiation
experienced by a space craft \citep{ssr}.  The BATSE Mass Model is a
GEANT3 Monte-Carlo simulation code which simulates the expected background rate from
cosmic diffuse $\gamma$-rays, atmospheric albedo $\gamma$-rays and
cosmic-ray interactions.  This model has been used to flat-field the
entire 9 year BATSE data archive \citep{bammpaper}.

Each day in the data set is then used to create an all-sky map using
the LIMBO (Likelihood Imaging Method for BATSE Occultation) code
developed at Southampton \citep{alicante}.
Daily images have now been generated based on all 9 years of data in 7 energy channels covering
the energy range 25-160 keV.  These images further enable the compilation of long
time scale all-sky maps.  The Earth's limb, as seen by BATSE, is
140$\,^{\circ}$ in diameter and hence creates image artefacts which
must be cleaned using a non-linear CLEAN algorithm.
The CLEAN algorithm operates by defining two maps, the raw map and the
cleaned map.  The cleaned map is of the same dimensions as the raw map
but is intially empty.  The algorithm then searches for the brightest
pixel in the raw map and generates the expected PSF for a source at
this position.  A specified fraction of the PSF is then subtracted
from the raw map and the total flux which is removed is convolved with
an ideal PSF and added to the cleaned map.  The algorithm then
searches for the next brightest pixel and iteratively removes sources
from the raw map and places them into the cleaned map.  A similar
CLEAN algorithm has been used in radio interferometry for many years.
A full description of the CLEAN algorithm is beyond the scope of this
article but is descriped in detail in \citet{basspaper}.  The cleaned map
is then searched for sources using the SExtractor software \citep{sextractor}.

\section{Survey Results and Spectra}
The longest map currently compiled and cleaned consists of 500 days of
data summed over 7 energy channels and is shown in
Fig~\ref{fig:clean_map}.  It clearly suffers from
2$\,^{\circ}$x 2$\,^{\circ}$ pixel size, which makes identification of
individual sources in crowded regions, such as the galactic centre, impossible.  This
represents only $\sim$15$\%$ of the available data set which is
currently being analysed.  The map yields a peak significance of
990$\sigma$, which corresponds to the position of the Crab Nebula.
The 5$\sigma$ sensitivity limit is $\sim$5mCrab \citep{basspaper}.

Two maps were produced for use with SExtractor; one contained only the
cleaned components and is seen in Fig~\ref{fig:clean_map} and the
other contains the cleaned components with the residual flux from the
raw map added back in, the latter being the more scientifically realistic.  Using SExtractor on the latter map produced
the source list shown in Table~\ref{tab:cat}.  In crowded regions of
the sky multiple sources are listed as candiates as their known
positions all fall within a single pixel.  The fluxes listed are the
total flux over the 500 days from that detection position normalised
to the Crab flux.  However, a number of sources in
Fig~\ref{fig:clean_map} were not detected by SExtractor.  These
missing sources were extracted by guiding SExtractor with the positions
generated in the cleaning algorithm and are listed in
Table~\ref{tab:cat2}.  These flux detections cannot be independently
measured and so should only be taken as indicatiing the possibility of
sources at these locations.  It should be noted however, that forcing
SExtractor to extract at a position where there is no source returns
a zero flux.  

\begin{figure}[t]
\centering
\includegraphics[width=0.9\linewidth, clip]{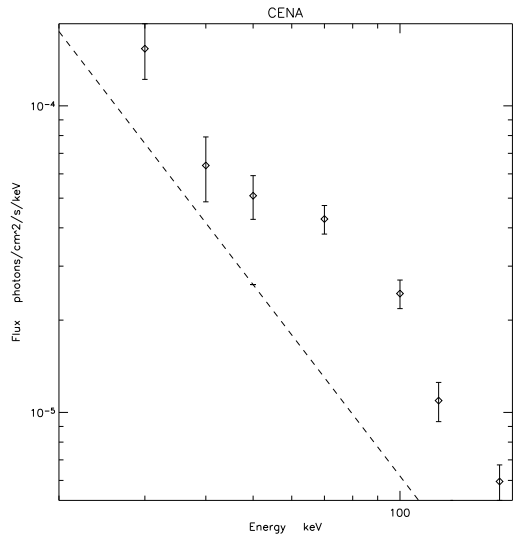}
\vspace{0.5cm}
\caption{Cen A time average energy spectrum of the period TJD
09448-09936. The dashed line represents a 10 mCrab source.\label{fig:cenA}}
\end{figure}

\begin{figure}[t]
\centering
\includegraphics[width=0.9\linewidth, clip]{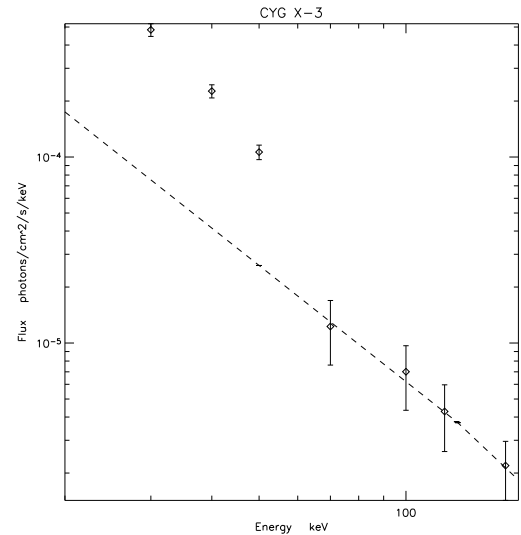}
\vspace{0.5cm}
\caption{Cyg X-3 time average energy spectrum of the period TJD
09448-09936.  The dashed line represents a 10 mCrab source.\label{fig:cygx-3}}
\end{figure}

Additionally, images of the same period of 500 days have been
generated in a number of different energy channels allowing energy
spectra to be extracted.  A sample of different source spectra is
shown in Figs.~\ref{fig:cenA}-~\ref{fig:cygx-3}, together with a 10
mCrab source spectrum.

In Fig~\ref{fig:clean_map}, M82 is detected at the level of 14 mCrab.
This was unexpected as previous missions have not observed such
emission of this level.  However, when the energy spectrum of M82 is
extracted the emission is principally from those energy channels above 70
keV.  The Mass Model is most accurate at this energy level and this
may indicate that the detection is real and not an artefact generated
by the cleaning process.  This should be resolved once the entire 9
year data set is imaged and cleaned.

\section{Discussion and Future Work}
The first true all-sky image of the $\gamma$-ray sky since HEAO1-A4 has
been produced in a single 25-160 keV energy band.  The all-sky map is
based upon  $\sim$500 days of data and shows the locations of $\sim$30
known $\gamma$-ray sources which are also listed in  Tables~\ref{tab:cat}~$\&$~\ref{tab:cat2}.
The diagonal size of one image pixel is 2.8$\,^{\circ}$, hence any
source which has a positional error less than this lies in the
expected pixel.  The vast majority of sources which we have associated
with our detections have positional errors less than 2.8$\,^{\circ}$,
in several cases there are multiple sources which can be easily
associated with a single pixel.  However, there are also a number of
sources, particularly ESO 198-24, NGC 7582 and Mkn 78 in Table~\ref{tab:cat2}, which lie much
further away than 2.8$\,^{\circ}$ from their expected position.  In
these cases the nearest known high energy source is listed.

In addition, it has been possible to generate
maps in 7 narrower  energy channels and hence extract time-averaged energy
spectra of the brighter sources.  Work is progressing to compile all 9
years of the BATSE data into a single map which should have a
5$\sigma$ sensitivity of $\sim$2 mCrab.  For any source brighter than
$\sim$6 mCrab in the overall cleaned map it should be possible to
extract out a time average energy spectrum from the maps constructed
in the 7 narrow energy bands.

The team at Marshall Space Flight Center (MSFC) recently produced a BATSE
earth occultation catalogue of low-energy $\gamma$-ray sources
\citep{batselc}.  This catalogue is based on the standard EOT on data
which has not been flat-fielded.  As a result sources are pre-selected
for measurement and the background is assumed to be smooth and fit by
a second order polynomial on a short section of data surrounding each
source occultation.  Once 9 year data set is imaged, cleaned
and a complete catalgue generated it will be possible
to perform a complete comparison with the MSFC catalogue.  It
is interesting to note that on the currently analysed sample of
data there are detections, such as in the vicinity of M82, which
are not seen by the MSFC standard EOT.

\begin{table*}
\begin{center}   \caption{List of sources detected by BATSE in the 25 - 160 keV band for the period spanning TJD 09448 - 09936.  The fluxes have been normalised to the Crab Nebula flux.}
\renewcommand{\arraystretch}{1.2}
\begin{tabular}[h]{clccc} \hline
Measured Position & Source Name & Known Position & Position Error&  BATSE Flux  \\
(l,b)             &             & (l,b)          & (degs)             &(mCrab)  \\ 
\hline
(71.9, 3.8)   & Cygnus X-1  & (71.34, 3.07)   &  0.9 & 1142  $\pm$ 18 \\

(-175.9, -5.9)& Crab        & (-175.44, -5.78)&  0.5 & 1000  $\pm$ 17 \\

(0.0, 6.9)   & GRO J1719-24 & (0.18, 7.02)   &  0.2  & 214  $\pm$ 9 \\
             & H 1705-250   & (-1.41, 9.06)  &  2.6  &   \\

(-12.0, 2.0) & 4U 1700-377  & (-12.2, 2.2)   &  0.3  & 209  $\pm$ 8 \\
             & GX 349+2     & (-10.90, 2.74) &  1.3  &           \\

(-2.0, 0.9)  & GRS 1734-292   & (-1.11, -1.41) & 1.0 & 201  $\pm$ 9 \\
             & H 1743-322   & (-2.87, -1.61) & 2.7   & \\

(0.0, 0.2)   & XTE J1748-288& (0.68, -0.22) &  0.8   & 180 $\pm$ 8 \\
             & 1E 1740.7-2942   & (-0.88, -0.11) & 0.9  & \\
             & GX 359+2  & (-0.43, 1.56)    &  1.4   &\\
	     & GX 3+1    & (2.29, 0.79)     &  2.4   &\\

(-16.0, 2.0) & OAO 1657-415  & (-15.64, 0.31)&  1.7 & 128 $\pm$ 7\\

(0.1, 24.0)  & Sco X-1    & (-0.90, 23.78)   &  0.9  & 114  $\pm$ 7 \\

(-21.5, -4.7)& GX 339-4   & (-21.06, -4.33)  &  0.6  & 93  $\pm$ 6 \\

(-96.3, 3.7) & Vela X-1   & (-96.94, 3.93)   &  0.7  & 83  $\pm$ 6 \\

(16.7, 1.4)  & GX 17+2    & (16.44, 1.28)    &  0.3  & 77  $\pm$ 8 \\

(122.0, -30.0) & Mkn 348 & ( 122.27, -30.91) & 0.9   & 51  $\pm$ 8 \\

(4.6, -1.4) & GX 5-1     & (5.07, -1.02)    &  0.6   & 46  $\pm$ 5 \\

(44.3, 0.0)  & GRS 1915+105 & (45.40, -0.23) &  1.1  & 43 $\pm$ 5 \\

(-2.00, 6.0) & GRO J1719-24 & (0.14, 6.99)   &  2.4  & 42  $\pm$ 5 \\

(-60.00, 0.0) & GX 301-2    & (-59.90, -0.04)&  0.1  & 42  $\pm$ 4 \\

(80.00, 0.6)  & Cygnus X-3  & (79.84, 0.69)  &  0.2  & 36  $\pm$ 5 \\

(-50.00, 19.5) & Centaurus A& (-50.48, 19.42)&  0.5  & 33  $\pm$ 5 \\

(-158.83, -4.0)& H 0614+091 & (-159.11, -3.38)&  0.7  & 31  $\pm$ 8 \\

(-24.00, 0.0) & H 1624-490 & (-25.08, -0.26)  &  1.1  & 17  $\pm$ 3 \\

(156.00, 76.0) & NGC 4151 & (155.08, 75.06) &  1.0    &16  $\pm$ 3 \\

(138.00, 42.0) & M 82     & (141.41, 40.57) &  2.9    &14  $\pm$ 3 \\

(-70.00, 66.0) & 3C 273   & (-70.05, 64.36) &  1.6    &12  $\pm$ 3 \\


\hline
\end{tabular}
\label{tab:cat}
\end{center}
\end{table*}

\begin{table*}
  \begin{center}   \caption{List of additional sources visible in
  Fig.~\ref{fig:clean_map} but not included in Table~\ref{tab:cat}.
  These sources were extracted using the locations where PSFs were
  generated in the CLEAN algorithm as an input to the SExtractor package.}
\renewcommand{\arraystretch}{1.2}
\begin{tabular}[h]{clccc} \hline
Measured Position & Source Name & Known Position & Position Error & BATSE Flux  \\
(l,b)             &             & (l,b)          & (degs)             &(mCrab)\\ 
\hline
 (-79.0, -59.0) & ESO 198-24 & (-88.36, -57.95) & 5.0      &  62 $\pm$ 8 \\
 (-5.0, -70.0)  & NGC 7582   & (-11.92, -65.70) & 5.1      & 55 $\pm$ 8 \\
 (146.1, 28.0)  & Mkn 78     & (151.10, 29.78)  & 5.0      & 40 $\pm$ 6 \\
 (-68.0, -38.0) & 4U 0357-74 &(-71.42, -37.29)  & 2.8      & 38 $\pm$ 6 \\
 (-47.9, -2.0)  & Circinus Galaxy  &(-48.67, -3.81) & 2.0 & 28 $\pm$ 5 \\
 (76.8, -2.0)   & EXO 2030+375  & (77.15, -1.24) & 0.8    & 22 $\pm$ 5 \\
\hline
\end{tabular}
\label{tab:cat2}
\end{center}
\end{table*}

\end{document}